\documentclass[acmlarge]{acmart}
\AtBeginDocument{%
  \providecommand\BibTeX{{%
    \normalfont B\kern-0.5em{\scshape i\kern-0.25em b}\kern-0.8em\TeX}}}

\setcopyright{acmlicensed}
\copyrightyear{2024}
\acmYear{2024}

\begin{document}

\title{Ephemeral Myographic Motion: Repurposing the Myo Armband to Control Disposable Pneumatic Sculptures}

\author{Celia Chen}
\authornote{Both authors contributed equally to this research.}
\email{clichen@umd.edu}
\orcid{0009-0008-9967-8968}
\author{Alex Leitch}
\authornotemark[1]
\email{aleitch1@umd.edu}
\affiliation{%
  \institution{University of Maryland}
  \streetaddress{4130 Campus Drive}
  \city{College Park}
  \state{Maryland}
  \country{USA}
  \postcode{20740}
}

\renewcommand{\shortauthors}{Chen \& Leitch}

\begin{abstract}
This paper details the development of an interactive sculpture built from deprecated hardware technology and intentionally decomposable, transient materials. We detail a case study of "Strain" - an emotive prototype that reclaims two orphaned digital artifacts to power a kinetic sculpture made of common disposable objects. We use the Myo, an abandoned myoelectric armband, in concert with the Programmable Air, a soft-robotics prototyping project, to manipulate a pneumatic bladder array constructed from condoms, bamboo skewers, and a small library of 3D printed PLA plastic connectors designed to work with these generic parts. The resulting sculpture achieves surprisingly organic actuation. The goal of this project is to produce several reusable components: software to resuscitate the Myo Armband, homeostasis software for the Programmable Air or equivalent pneumatic projects, and a library of easily-printed parts that will work with generic bamboo disposables for sculptural prototyping. This project works to develop usable, repeatable engineering by applying it to a slightly whimsical object that promotes a strong emotional response in its audience. Through this, we transform the disposable into the sustainable. In this paper, we reflect on project-based insights into rescuing and revitalizing abandoned consumer electronics for future works.
\end{abstract}




\received{February 28, 2024}

\maketitle

\section{Introduction} 
"Strain" is an emotive sculpture that explores “unmaking,” disposability and reuse by resurrecting two orphaned digital artifacts and using them to control an organic, reactive pneumatic sculpture made of common disposable objects. Through doing this, we explore the relationship of disposability as it applies to mass-produced small objects and limited-production HCI devices. The sculpture is designed to bulge, lean, shrink, and sag as an active practice, suggested by Song and Paulos [10], even as it retains some elements of rigid form over its intentionally-short lifetime.

As novel devices waft through the consumer marketplace, technologies that still work as described are often discarded well before their latent interaction capabilities are fully explored. Interactive artworks are a unique output target for HCI inputs that appear to have wound up their productive life. Rather than targeting commercialization, an artwork provides a useful set of limited exploratory goals for any new technology. At present, HCI object creation is a field that promotes quick iteration and equally quick disposal. This disposal cycle accelerates e-waste production, while making it seem as though there is no space for emergent use of devices that did not initially find commercial success. Rather than let dead devices lie, we would like to explore postmortem open hardware possibilities, and examine how closed devices might be able to find a longer life through hardware reversal or software support.

Strain is a kinetic sculpture based on an array of soft pneumatic actuators that are puppeteered using muscle signals detected using a discontinued myoelectric armband - the Myo from Thalmic Labs - worn by the user. Constructed from condoms, bamboo skewers, and custom 3D-printed PLA parts, the piece plays with the “disposable” aesthetic present in most DIY rapid-prototyping to explore what it means to work with trash; both in the form of HCI devices that were commercially unsuccessful, and everyday objects that are made to be thrown away. Strain examines how daily-use materials produced by mass industrial standardization can be depended upon to have a certain degree of consistency - enough to repurpose into small robots or sculptural interventions, especially with the help of FDM printing. 

What alternative technological roles, relationships, and aesthetics are available when we take small parts seriously, as our daily access to consistent and precise objects expands? What space does that consistency allow for exploration of other types of disposable precision?  

Our work builds upon the emerging concept of 'unmaking' in HCI, which focuses on the creative potential of destruction, decay, and deformation of physical artifacts [10]. Unmaking challenges conventional notions of obsolescence and waste by reframing disposal as an ongoing process of transformation that can imbue objects with new meaning and value. By intentionally designing for the unmaking of our sculpture through the use of biodegradable and transient materials, we explore how this approach can promote more sustainable and emotionally resonant relationships with interactive artifacts.

\section{Revitalizing Abandonware}
The Myo armband by Thalmic Labs is an infamous example of an extraordinarily innovative device undercut by challenging commercial realities. First introduced in 2014, this wearable pioneered subtle gesture control through a combination of electromyography (EMG) and inertial sensors [1]. Thin stainless bands wrapped around the forearm to press eight clinical-grade EMG contact points against key muscles controlling hand motions. Detecting the electrical spikes of flexing muscle groups enabled the device to discern fine finger poses. This combined with accelerometers that tracked overall arm orientations as users gestured, waved, pointed, or clicked their virtual hands.

Reviewers praised the Myo’s surprisingly robust hand-decoding algorithms and latched onto its promising applications for accessibility, rehabilitative biofeedback, and intuitive device operation [3]. Yet just four years post-release, Thalmic abandoned the armband as advances in optical tracking and changes in company direction left the product unsustainable [7]. Despite its impressive sensing capabilities, the Myo armband fell victim to the iterative nature of consumer technology design, which often inadvertently creates breakdowns and obsolescence [5] . While planned obsolescence pre-2004 was arguably an industrial policy intended to accelerate development in the ICT industry, it is now a more commercially-focused business plan. The iterative, market-driven design processes behind modern consumer technology frequently create unintentional forms of obsolescence, driven more by software constraints and commercial priorities rather than hardware limitations [5]. As a commercial device, the Myo simply turned out to not be as effective at gesture detection as cameras with depth sensors and LIDAR. As a non-invasive device, it cannot detect muscle contractions and label them consistently enough to be useful in more than research cases, and it also lacks the reliable, familiar consistency of a joystick or a button. The defined feature set and lack of continued software updates for the Myo armband enables a helpfully bounded exploration of its capabilities. With consistent gesture detection across a limited range, the Myo provides a robust package for basic sensing and input rather than complex control.

The Programmable Air, from Amitabh “tinkrmind” Shrivastava, represents a contrasting niche open hardware project purposefully constrained in scope.This pneumatic control kit, released at a small scale in 2019, is designed to help people familiar with the Arduino platform prototype small-scale soft robotics by combining a series of 12V pumps, solenoid valves, and an integrated pressure sensor [9]. The Programmable Air allows precision actuation well-suited for inflating balloon-based prototypes. Its elegant simplicity arises from a narrow focus, which divorces it from mass production constraints. Unlike the Myo, which was discontinued as a failure because it did not successfully locate a mass market, the Programmable Air is a popular success in a very small niche: a tool that can be used, programmed, and updated openly by the people most interested in prototyping soft robots. Figure 1 depicts the components and data flow in the Strain system.

\begin{figure}[h]
  \centering
  \includegraphics[width=0.6\linewidth]{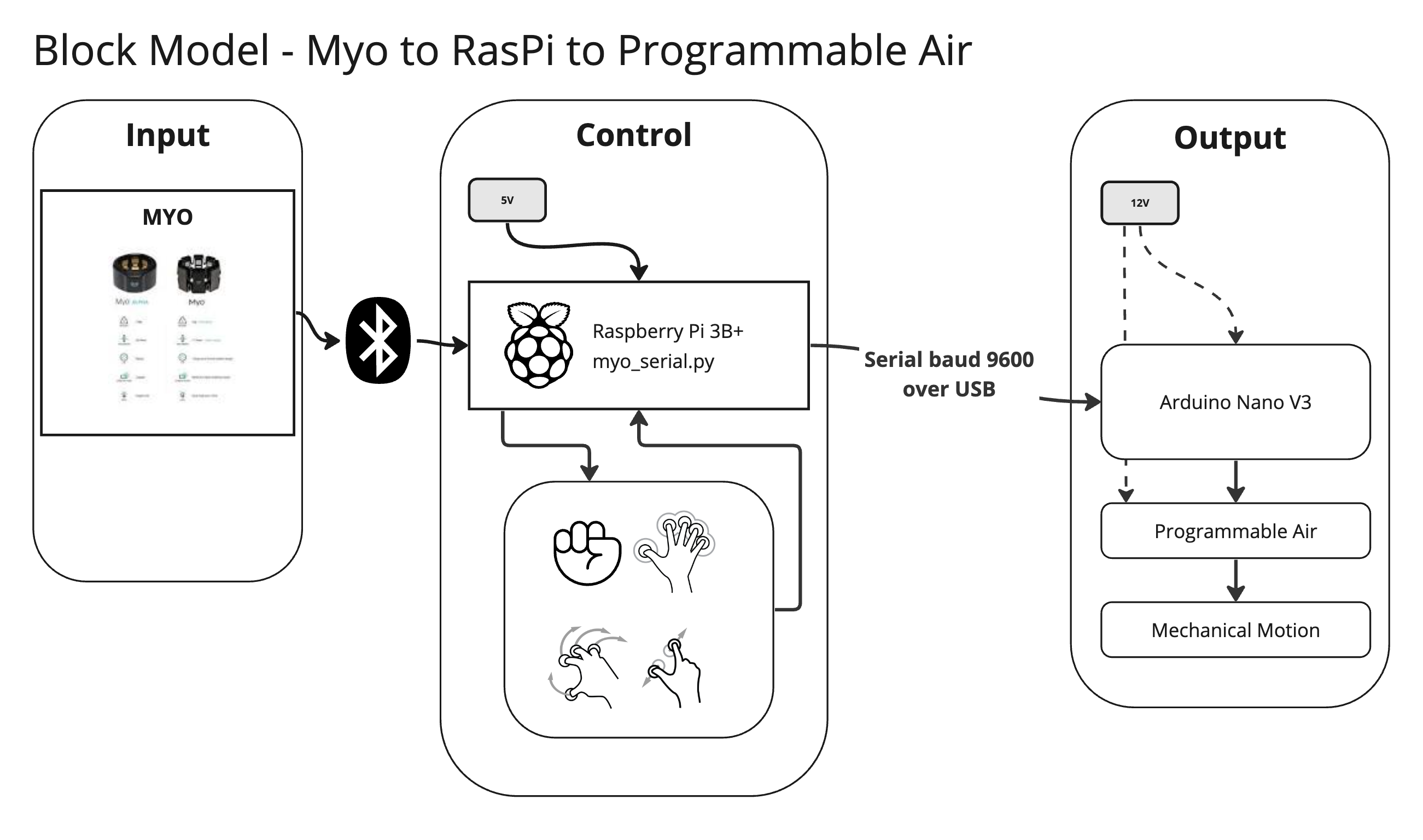}
  \caption{This diagram depicts the block model for Strain. The Myo armband detects hand gestures from the user. These are sent to the Raspberry Pi, which interprets the gestures and transmits corresponding signals to the Arduino Nano. The Arduino then actuates the pneumatic outputs to manipulate the sculpture.}
\end{figure}

The two devices stand in sharp contrast and relation to one another: one simple, very effective, and successful only in limited-release, the other a sophisticated and highly finished sensor packet designed for mass production that died before finding its audience. The success of Strain as a technical project, then, is one of maintenance - repairing only the broken parts, and salvaging those that worked all along.

\section{Sculpting with Transience}
Materials selection plays a key role in embracing impermanence within interactive sculptures [2]. As the authors state, "Soft robots can elegantly adapt to the huge variety of shapes and sizes that people come in...[they are] inherently soft, and also therefore more adaptable." Strain embraces the aesthetics of impermanence and unmaking through its use of decomposable materials like latex, bamboo, and PLA plastic. These materials are designed to break down over time, allowing the pneumatic bladders to wear out over the course of time. The “active” parts of the sculpture, the bladders, can then be replaced or allowed to decay. This unmaking process is not a flaw or failure, but rather an intentional part of the artwork's lifecycle that invites reflection on the transience of physical objects. The sculpture’s activation makes the actuators work by organically warping through several of the recognized forms of the “unmaking” language [10]. This recovery can happen because of the properties of the latex itself - organized to have a limited range of motion, it does not slip back through the minimized PLA shackles. 

Strain uses natural latex condoms as its central, playful pneumatic bladders to actuate the sculpture motion and  animate the piece. Natural latex condoms are generally produced to be much thinner but stiffer than comparable latex balloons.This makes them more sensitive to air-pressure changes internally, as well as being stronger and more leak-resistant. They are therefore surprisingly cheap, reliable and repeatable for prototyping soft robotic actuation, especially compared to alternatives such as cast silicone. Silicone tends to be messy, and requires a custom fabrication space dedicated to degassing and handling liquid plastics. Not all silicones are skin-safe through all parts of their lifecycle. Condoms are, by contrast, broadly considered very skin-safe so long as no-one is allergic to their natural latex. Reliable condom brands test their products extensively against leakage or breakage, as people can contract serious illness if they fail. Their adaptability accommodates repeated shape deformations from air pressure changes, as demonstrated through air burst testing simulating use conditions [4]. Figures 2 and 3 show Strain in its inflated and deflated states respectively.

\begin{figure}[h!]
\begin{minipage}{0.45\textwidth}
\centering
\includegraphics[width=\textwidth, angle=270]{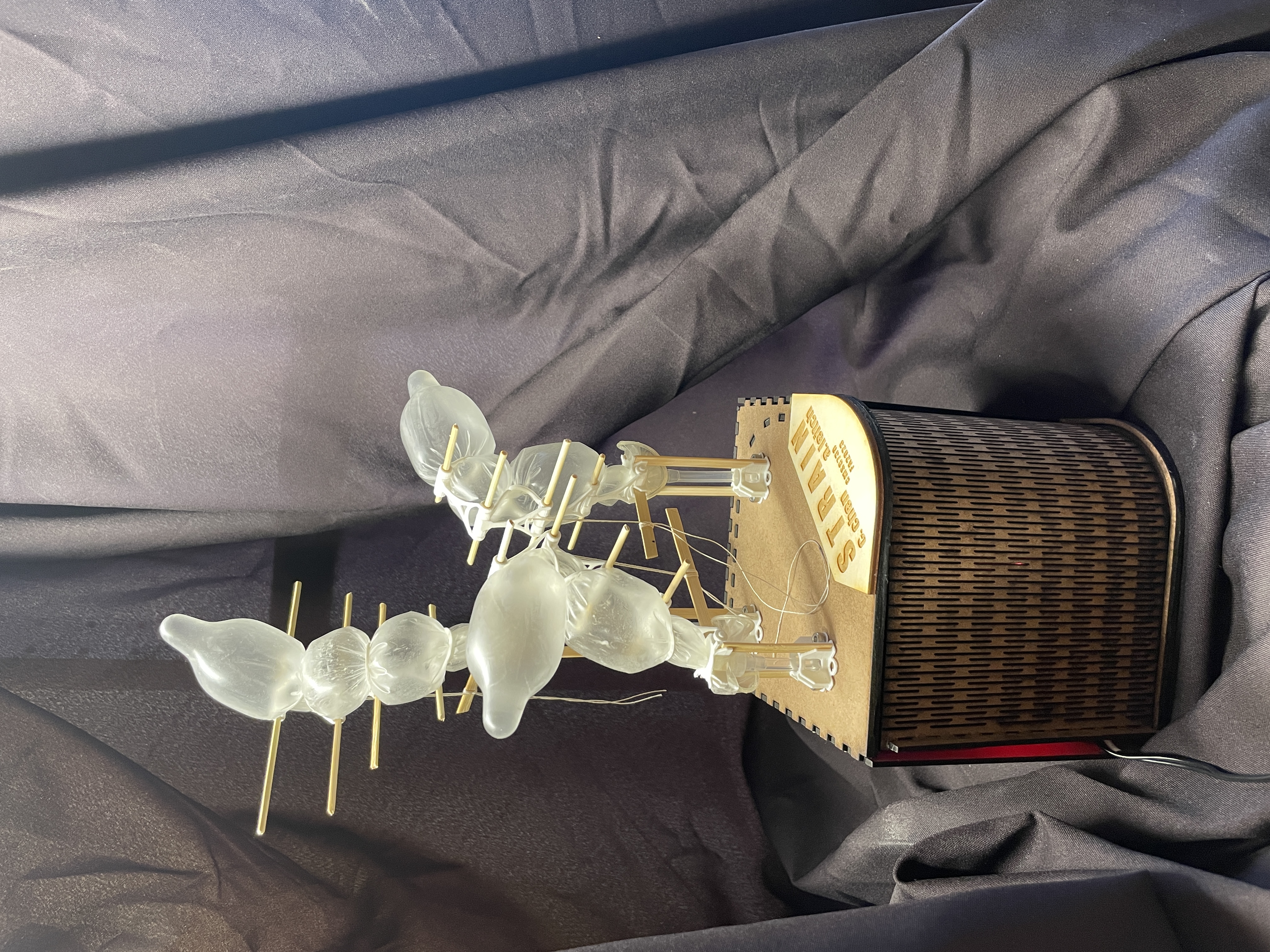}
\caption{Strain inflated. The outer condoms fill with air, actuating their structures outward. The central condom inflates inward, straightening the flexible PLA "neck" of the sculpture.}  
\end{minipage}
\hfill
\begin{minipage}{0.45\textwidth}
\centering 
\includegraphics[width=\textwidth, angle = 270]{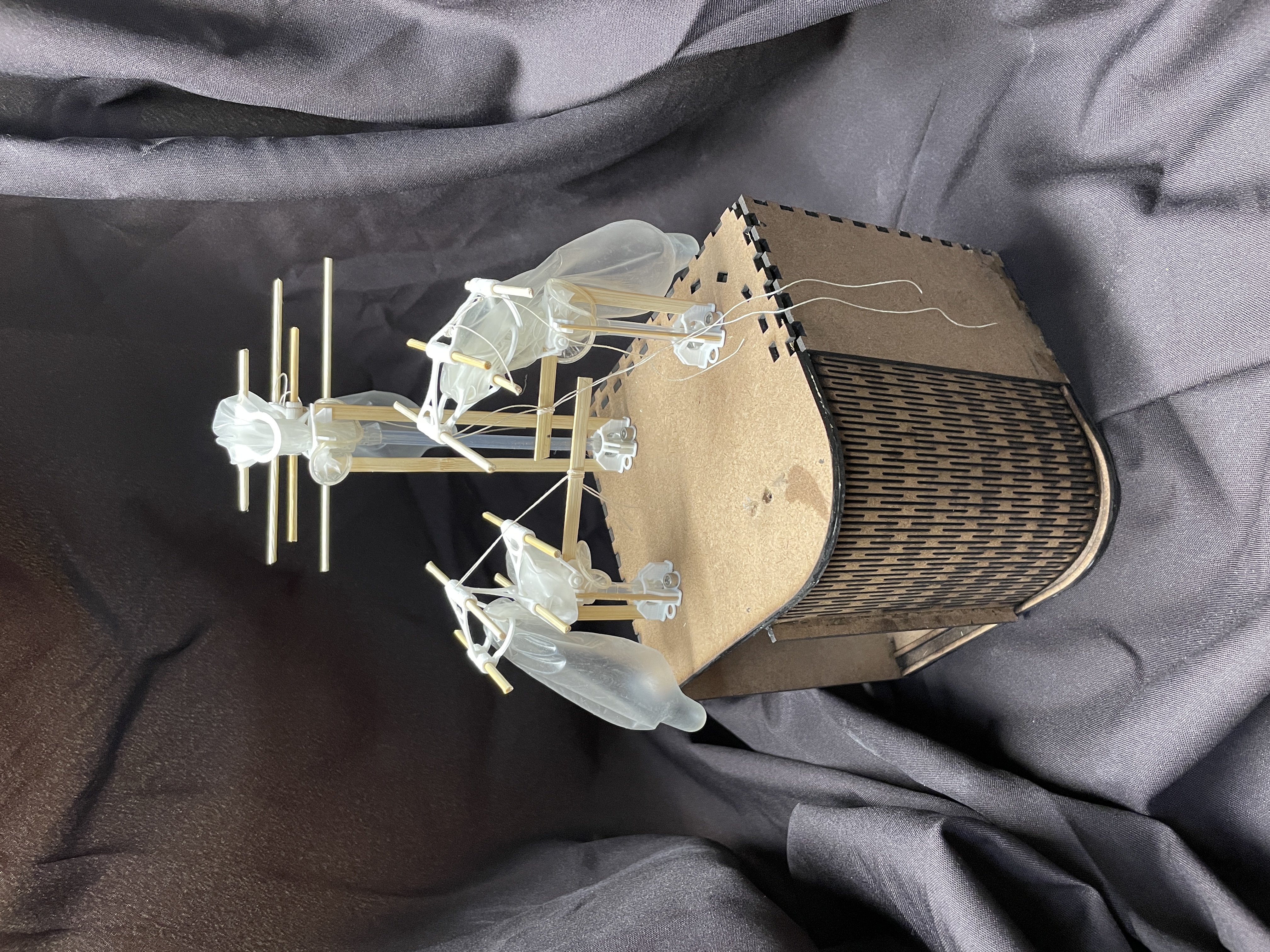}
\caption{Strain deflated. Without internal air pressure, the outer condoms relax, allowing the arm structures to fold inward. Simultaneously, the deflated central bladder bends backwards.}
\end{minipage}
\end{figure}

To build the superstructure around the condoms, we used PLA plastic that was 3D printed to resemble a series of interlinked “wishbones” and “shackles” with bamboo skewers as the pivot for the chain links. Bamboo’s naturally-occuring biosilica makes it very laterally strong - much stronger than plastics - which is useful in pneumatic work given the large mechanical advantage pneumatics have. It is simply not likely to snap or wear out after even hundreds of repeated motions of the condom actuators. Bamboo is also very light for that strength, which allows for much more delicate motions than heavier metal pivots, and much tougher than wood disposables. Bamboo famously grows and decomposes quickly and reliably, making it an ideal candidate for sculptures that aren’t built to last [6]. Assessing bamboo’s mechanical properties reveals remarkable tensile strength and rotational durability ideal for compliant interactive joints [8]. In addition, bamboo is broadly available in many sizes of skewer and stir-stick, generally at consistent sizes that are easy to find. Basing our PLA shapes around available bamboo disposables allowed us to put together a small library of parts that would work well to explore many possible volumetric shapes.

Design and fabrication of the “spine” piece required custom modular attachments. These were based around both organic/artistic concerns and constraints of repeatability. Using Autodesk Fusion 360, we modeled “wishbone” and “shackle” clips to link the pneumatic spine pieces into a flexible chain, through which we threaded the pneumatic bladder. We printed on a tuned Prusa i3 MK3S, using PLA filament. In order to make the sculpture with the least machine time and PLA filament possible, we designed the parts to print flat in place from the print bed, with no supports. This was intended to minimize iteration waste, and maximize how much work we could get from limited machine time. All but one part for the project prints within 3 minutes, with larger parts constrained to no more than 10 minutes per piece - enough that we could very quickly work out how they would impact the final sculpture. PLA was chosen as it is at least in theory biodegradable at high heat through industrial composting. In practice, this type of composting is less accessible, and it might be preferable to build the forms through more traditional means of working with bamboo, such as steam-forming. As it is, the goal is to minimize PLA trash through small iterations in readily available single-material prints.

\begin{figure}[h]
  \centering
  \includegraphics[width=0.25\linewidth, angle=90]{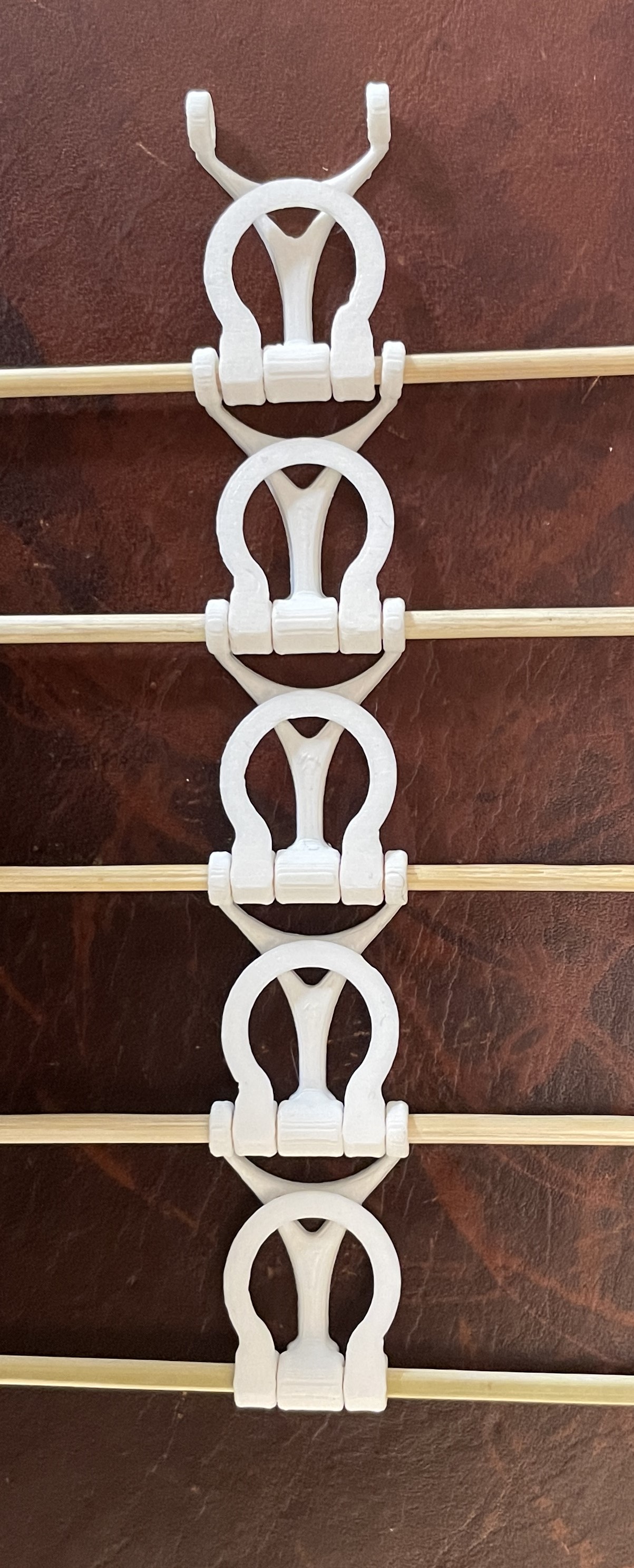}
  \caption{This image shows some of the modular 3D printed clips used to assemble the actuator chain spine. The interlocking shape enables flexible arrangement for different volumes.}
\end{figure}

Using small, fast-printing, modular FDM components with theoretically degradable mass-produced rigid bodies arranged into a volumetric structure allowed for quick sculptural iteration. These disposable elements then are assembled into a finished object coordinated by our rescued control hardware. By incorporating decomposable components, this process offers opportunities to challenge the accumulation of electronics as e-waste. Allowing parts to degrade over time could help refocus attention on product life cycles that are often obscured in commercial spaces, emphasizing continual iteration speed and technological advancement.

\section{Making New From Exploring The Old}
Out of this research project, we have retained a small library of components and resources that can be published and shared to the community. Strain has been useful for assembling a small parts library of 3D printed clips and spine components for further pneumatic explorations. Each piece can print quickly and make efficient use of limited resources. 

The project also has led to two software updates. We created a new Arduino sketch with homeostasis-maintenance code to keep the sculpture softly inflated and “moving”. We also developed a Raspberry Pi script that enables communication between the Myo and an Arduino using serial data transfer, which built on existing Python code for controlling robots that was not originally compatible. Together, these updates allowed us to resuscitate the Myo as a controller for interactive artwork. 

The creation of reusable hardware and software resources shows how this type of research can be productive, especially as we discover new ways to openly share interventions that specialized devices could leverage to maintain utility over time.

\section{Conclusion}
Through Strain, we demonstrate how the concept of unmaking can be applied to the design of interactive sculptures that evolve and transform over time. By foregrounding the creative and theatrical potential of bulge and sag, coupled to intentionally-fragile everyday materials, our work contributes to the growing dialogue around unmaking as a sustainable and emotionally rich approach to HCI design. We hope that our exploration of intentionally degradable and unmake-able artifacts will inspire further research into this emerging area.

In working through this case study of reanimating HCI artifacts, we highlight interesting tensions at the intersection between technology lifecycles, planned obsolescence, and concepts of sustainability. By working with the latent interactive potential in a myoelectric device that might seem destined to collect dust, we raise questions around the criteria that determine ultimate usefulness amidst the relentless product iteration of the technology sector.

Likewise, realizing the Myo's decade-old promise of gesture control through a niche pneumatic prototype platform leads to reevaluation of the solutions that dominate technical production, as specialized passion projects make space for depth over scale.

Strain blends this technological pairing with biodegradable components to invite a perspective on electronics as transient rather than enduring. Consumer grade artifacts do not need to persist indefinitely; their components can return to the environment through intentional design. By weaving transience into interactive sculptures, we explore how technological planned obsolescence can give way to planned biodegradation by design. Amidst rampant commercial iteration, embracing ephemerality allows us to help reduce accumulative waste.

\begin{acks}
{We would like to thank Dr. Huaishu Peng and Zeyu Yan from the University of Maryland's smARTlab for their guidance and support through the development of this project for the CMSC730 Human-Computer Interaction course. Their insights greatly assisted our design and implementation. We also thank Amitabh Shrivastava, inventor of the Programmable Air, for providing the hardware, as well as Kari Love for her expertise in soft robotics and the Programmable Air specifically. We would like to thank the membership of NYC Resistor for their technical support during the development of this project in 2023, and the attendees of NYU’s ITPCamp 2023 for their feedback on the sculpture-in-progress. Finally, we want to acknowledge the Open Source Hardware Association (OSHWA) for their work in supporting open technologies like the Programmable Air through community building and open source certification.}
\end{acks}

\bibliographystyle{ACM-Reference-Format}
\bibliography{bibliography}
[1] Evan Ackerman. 2014. Myo Armband Provides Effortless Gesture Control of Robots, Anything Else. IEEE Spectrum (February 11, 2014). https://spectrum.ieee.org/myo-armband-provides-effortless-gesture-control-of-robots-anything-else\\

\noindent [2] Matthew Borgatti, Kari Love. 2019. Make: Soft Robotics (Maker Media, Nov 2019).\\

\noindent [3] Lisa Eadicicco. 2016. This Futuristic Armband Lets You Control Your Computer Like Magic. Time Magazine (Jan 20, 2016). https://time.com/4183108/thalmic-labs-myo-armband-review/\\

\noindent [4] MJ Free, EW Skiens, and MM Morrow. 1980. Relationship Between Condom Strength and Failure During Use. Contraception 22, 1 (July 1980). DOI:https://doi.org/10.1016/0010-7824(80)90115-8\\

\noindent [5] Steven J. Jackson and Laewoo Kang. 2014. Breakdown, obsolescence and reuse: HCI and the art of repair. In CHI '14: Proceedings of the SIGCHI Conference on Human Factors in Computing Systems, 449–458. https://doi.org/10.1145/2556288.2557332\\

\noindent [6] Walter Liese and Michael Köhl. Bamboo: The Plant and its Uses. Springer International Publishing, 2015.\\

\noindent [7] Lucas Matney. 2019. CTRL-labs scoops up Myo armband tech from North. TechCrunch (June 27, 2019).
https://techcrunch.com/2019/06/27/ctrl-labs-scoops-up-myo-armband-tech-from-north/\\

\noindent [8] M.N. Norizan et al., “Mechanical Performance Evaluation of Bamboo Fibre Reinforced Polymer Composites and its Applications: A Review,” IOP Conf. Ser. Mater. Sci. Eng., vol. 1163, no. 1, p. 012029, 2022.\\

\noindent [9] Amitabh Shrivastava. 2023. About Us. Programmable Air. (2023). https://www.programmableair.com/about-us  [Accessed: 14- Feb- 2023]\\

\noindent [10] Katherine W. Song and Eric Paulos. 2021. Unmaking: Enabling and Celebrating the Creative Material of Failure, Destruction, Decay, and Deformation. In CHI Conference on Human Factors in Computing Systems (CHI '21), May 8–13, 2021, Yokohama, Japan. ACM, New York, NY, USA, 12 pages. https://doi.org/10.1145/3411764.3445529\\

\appendix \section{Appendix}
\noindent Additional documentation and media related to this project are available at the Hackster.io page: https://www.hackster.io/\\cmsc730-group-7/strain-6ee28a
\end{document}